\documentstyle[psfig]{l-aa}
\begin{document}

%


\def \etal         {{\rm et~al.}~}
\def \kms          {\mbox{ km\,s$^{-1}$}}
\def \Msun         {\hbox{$\rm M_{\sun}$}}           
\def \Lsun         {\hbox{$\rm L_{\sun}$}}           
\def \mpc          { \mbox{Mpc}}
\def \kpc          { \mbox{kpc}}
\def \zero	   {$_\circ$}
\def \re	   {R_{\rm e}}

   \thesaurus{11         
              (11.06.2;  
               11.07.1;  
               11.11.1;  
               11.19.6)} 

   \title{ Multi-component models for disk galaxies}

   \subtitle{I. Stellar rotation and anisotropy}

   \author{Ezio Pignatelli \and Giuseppe Galletta }

   \offprints{G.~Galletta}

   \institute{Dipartimento di Astronomia, Universit\`a di Padova, \\
              Vicolo dell' Osservatorio, 5, I--35122 Padova, Italy. }

   \date{ Received; accepted }

   \maketitle
 
   \markboth{E. Pignatelli \& G. Galletta: Multi-component models 
  	     for disk galaxies. I}{E. Pignatelli \& G. Galletta: 
             Multi component models for disk galaxies. I}


   \begin{abstract}
   
   We present here a self-consistent, tridimensional model of a disc
   galaxy composed by a number of ellipsoidal distributions of matter 
   having different flattening and density profile. The model is 
   self-consistent and takes into account the observed luminosity 
   distribution, the flattening profile and the stellar rotation- and
   velocity dispersion- curves. In this paper we considered the particular 
   case of a disc galaxy composed by two spheroidal bodies: an exponential 
   disc and a bulge following the $r^{1/4}$ law. 
     
   We studied the behavior of the stellar rotation- and velocity 
   dispersion- profiles along the sequence of S0s and 
   Spirals, identified by an increasing disc-to-bulge ratio. Inside every 
   class, kinematic curves were produced by changing the relative 
   concentration of the two components and the inclination of the galaxy with 
   respect to the line of sight. The comparison with observational data 
   requires only two scaling factors: the total mass of the galaxy, and the 
   effective radius.

   The model allows also to detect the presence of anisotropy
   in the velocity distribution. In the special case of S0s, we explored 
   the sensitivity of the kinematics of the model 
   by changing the anisotropy and 
   the flattening of the bulge. For intermediate flattening 
   ($0.4\le b/a \le 0.85$) 
   it is possible to distinguish a change of anisotropy of 15\% 
  
   To show a real case, the model has been applied to the photometric and 
   kinematic data of NGC 5866. We plan to apply these models to a
   larger database of S0 galaxies in a future paper.
 

   \end{abstract}

\section{Introduction.}\label{Intro}

The study of the light distribution and the {\it stellar} kinematics in 
disk galaxies is an important tool to understand their structure. 
It is the only way to study stellar systems which are gas poor such as the 
S0s, and may be crucial to distinguish between different scenarios of galaxy
formation. It allows to discuss if a bulge is flattened by rotation or
by anisotropic residual velocities and how the different components of
a galaxy mutually interact.

In the past, in front of the success achieved in studying the {\it gas}
kinematics of disk galaxies, much less attention has been paid to the
modeling of their {\it stellar} kinematics from the observations. A
reason of this was the lower extension of the available stellar
rotation curves, based on absorption lines, with respect to that
obtained with emission lines (optical or 21cm). However, the inner
part of the galaxies is often poor of cold gas, and can be studied
with data from the stellar component only. In addition, the 
progresses in detectors and data analysis techniques (Sargent et al. 1977, 
Bertola \etal\ 1984, Kuijken \& Merrifield 1993) extended
the range explored with stellar data and produced a large quantity
of stellar rotation- and velocity dispersion- curves, making compulsory
the creation of realistic theoretical models for their interpretation.

However, the modeling of stellar components in disk galaxies must
overcome several difficulties. First, the non-negligible thickness of
stellar disks, compared to the gas ones, that requires a
tridimensional approach. Second, the fact that the observed rotation
curve does not represent the circular rotation defined by the
potential, because of the partial transparency of the stellar body and
of the presence of the velocity dispersion. The stellar rotation
curves are integrated along the line-of-sight across regions of
different kinematics and are influenced by the gradient of mass of
these regions. As consequence, the line-of-sight velocity distribution 
present deviation from the classical, Gaussian shape. The influence of the 
velocity dispersion and its eventual anisotropy are particularly important 
in early-type galaxies, where the velocity dispersion is of the same order 
of magnitude of the rotational velocity. This last effect is known with the 
name of {\it asymmetric drift} and the first attempt to evaluate it from 
observations date back to 1961, with the van der Hulst's work on the 
observations of NGC 4111 made by Humason and Oort (Van der Hulst 1961). 
More recently, the influence of the asymmetric drift on the stellar rotation 
curves has been taken into account by some authors (\cite{BC},
Kormendy 1984, Illingworth \& Schechter 1982, Fillmore \etal 1986,
Zeilinger \etal 1990). These theoretical models and 
interpretations of stellar kinematics were assuming an isotropic
velocity dispersion or were limited to edge-on galaxies. Most part of
models were simulating the galaxy by means of the disk component only,
with the only exceptions of the Fillmore \etal\ (1986) model and the
Zeilinger \etal\ (1990) work.  Both papers were based on non
self-consistent models and the first one made use of photometric
data only. 

In the last few years, a number of self-consistent, multi-components
dynamical models for early-type disk galaxies have been presented 
(e.g. \cite{cretton99}, \cite{emsellem99}; see also Merritt 1999 for a review).
Some of them are based on the Multi-Gaussian Expansion approach
(\cite{emsellem94}, \cite{loyer98}) for inferring the mass
distribution from the observed luminosity; most of them adopt the
orbit-superposition method (\cite{schwarzschild79}, \cite{emsellem94},
Rix \etal 1997, \cite{vdm98}, Cretton \& van den Bosch 1999, \cite{emsellem99}) in
order to derive a velocity distribution able to reproduce the
observed kinematics (including the higher-order moments).
These works have established the importance of taking into account the
effects of the asymmetric drift, the projection along the line-of-sight
and the deviation from the pure Gaussian shape of the line-of-sight
velocity distribution due to the superposition of the different
components with different kinematic behaviors. 

While this last approach remains the most general and flexible way of
deducing the dynamical properties of a single galaxy from the observed
photometry and stellar kinematics, its application to a large database
of galaxies - in order to derive general relations between physical
parameters - is not straightforward, due to the large number of
parameters involved. A model with few parameters to be constrained by
the observations could be a better choice when dealing with a large
number of objects.

We present here a self-consistent, multi-component model of disk
galaxies based on an analytic approach to the problem. We also try to
derive the relations between the observed properties and the model
parameters that can be useful to the observers in interpreting the data of
a single galaxy. The real case of an S0 galaxy (NGC~5866) is also presented 
and modeled. 

\section{ The hypothesis of the model }\label{hypo}

\subsection{ The density profiles }

We want to describe a stellar system similar to a real one,
and in particular to a disk galaxy. A first step is to define a
suitable distribution function $f(x,v)$ containing the tridimensional
structure of the galaxy and its velocity field.

\begin{figure*}[ht]
\psfig{file=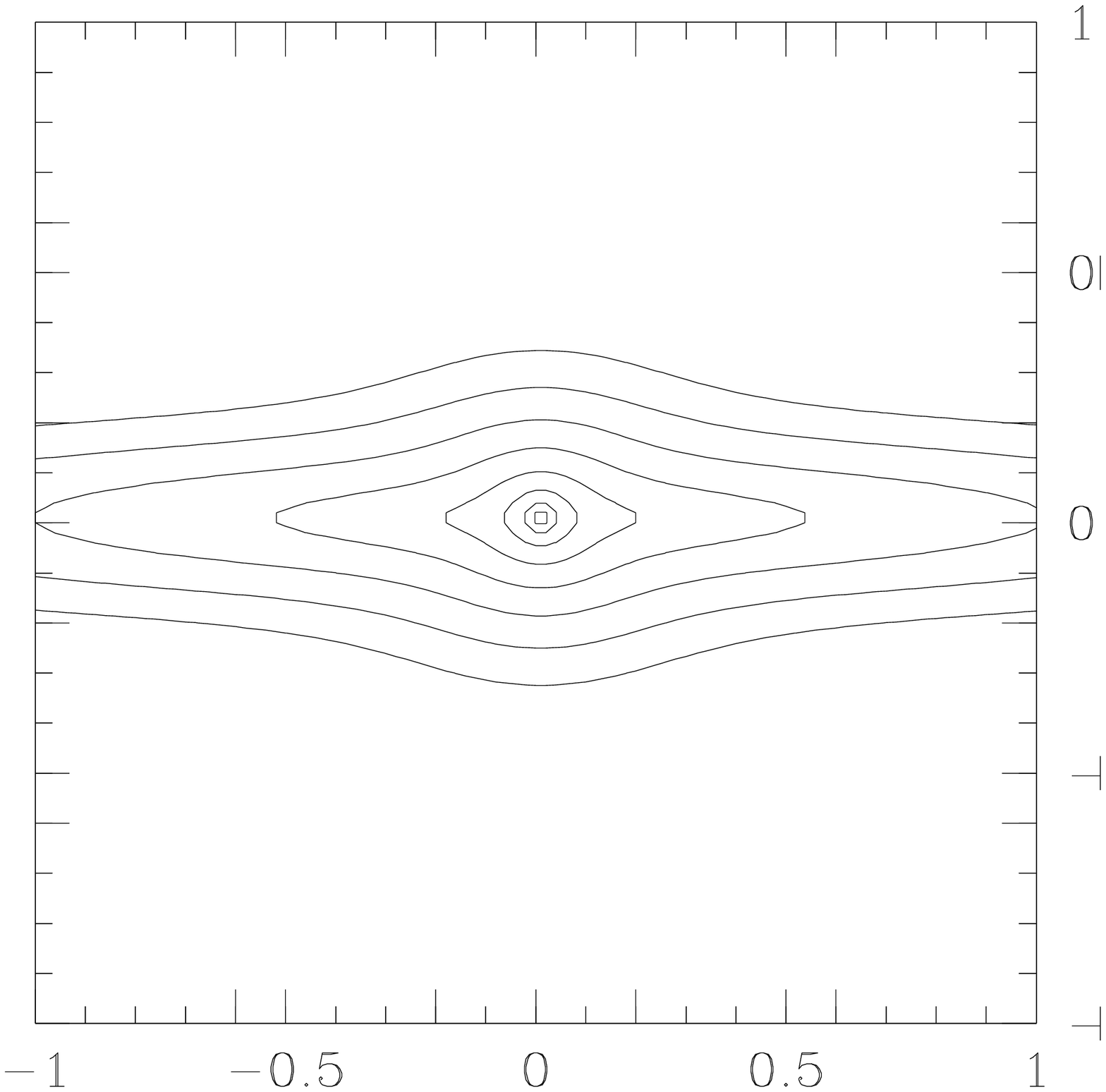,width=7.5cm}
\vspace*{-7.5truecm}
\hspace*{8.5truecm}
\psfig{file=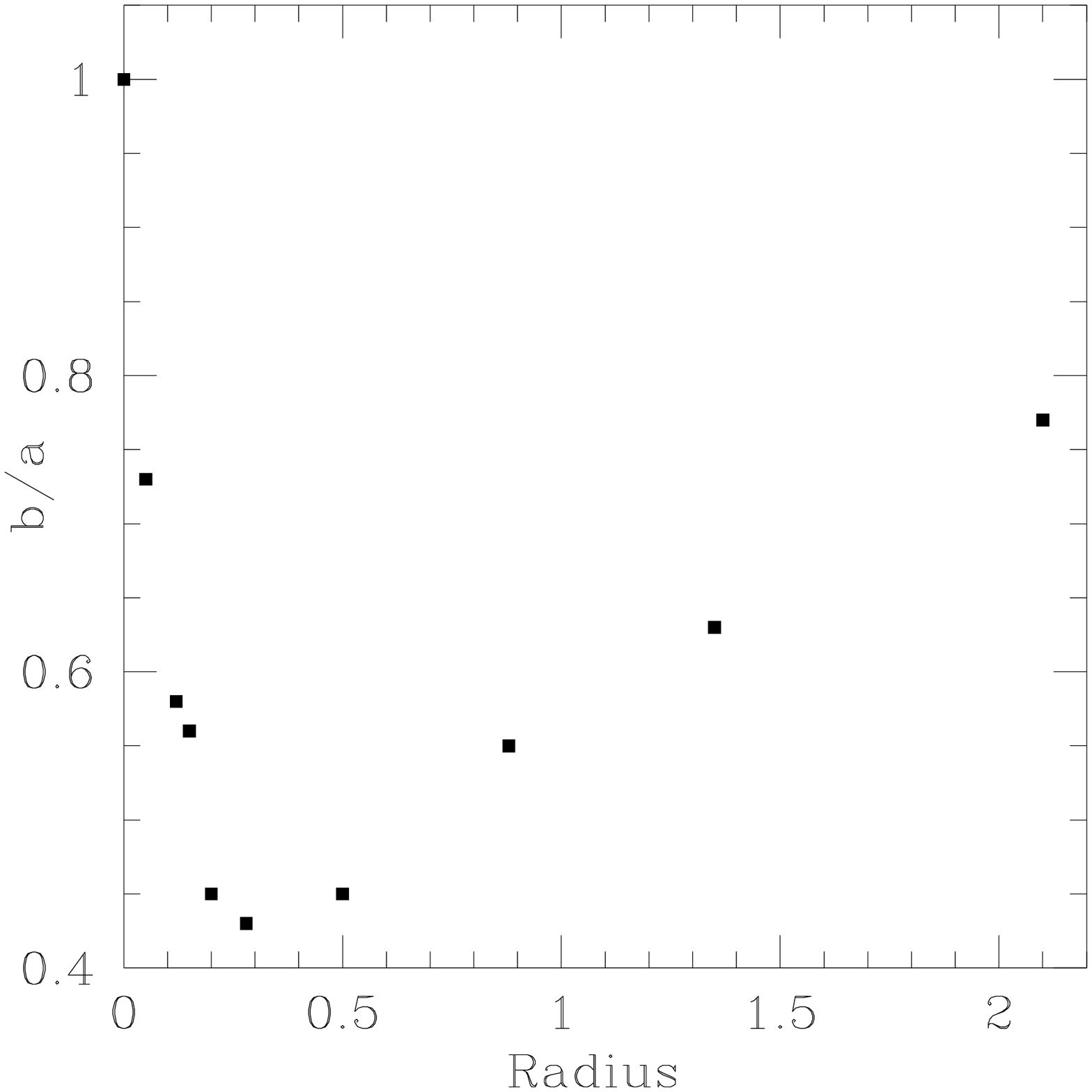,width=7.5cm}
\caption[]{{\bf Left:} Isophotes of a model composed by a spherical bulge
following the $R^{1/4}$ law, and an exponential disc with $b/a=0.1$.
{\bf Right:} The global axial ratio of the model shown in
\protect\ref{fig:isophotes} as function of the radius. 
Note the peak of isophotes flattening at intermediate radii,
obtained with two components of constant flattening.}
\label{fig:isophotes}
\end{figure*}

The structure of the disk galaxies can be modeled by the mixing of
different components: a bulge, a disk and eventually other components 
(lens, nuclear disk, halo). Every single component may be reproduced by an 
ellipsoidal density distribution whose projection on the sky generates 
elliptical shapes. If the axial ratio of each one of these components is 
constant with radius, but different for every component, we can apply a
generalization of the Newton's theorems for spherical bodies, that
makes simpler the calculation of the kinematic properties.  This
approximation allows anyhow to generate projected light distributions
with variable ellipticity, as observed. This is due to the
predominance of the different components at different distances from
the nucleus.  An example of a composite light distribution - which is
rounder in the center and in the outer regions, while flattens at
intermediate radii - is given in Fig.~\ref{fig:isophotes}. Spindle
isophotes may also be reproduced in this way.

Another choice concerns the shape of the different components.
Differently from the studies on the elliptical galaxies, the works
made on the shape of the disks reveals slightly or absent deviations
from the oblateness (Sandage \etal\ 1970, Grosb\"ol 1985, Binney \& de
Vaucouleurs 1981, Magrelli \etal\ 1992). The mean deviation suggested
should be $\le 5\%$. A different situation is found in some bulges,
whose structure may range from slight triaxial (Bertola \etal\ 1991)
to strongly triaxial in rare cases (Stark 1977).  

Taking into account all these facts, we decided to assume a cylindric
symmetry for the galaxy's components. This seems to be a good
compromise between a ``realistic'' description of a single galaxy, yet
with a large number of free parameters, and a general deduction of the
galaxy dynamics and structure with the minimum number of constraint
obtained from the observations.

Given the previous hypotheses, the density profile of each component
can be described in terms of the radial coordinate only, with the values
of their parameters constrained by observations.

As shown by Young (1976), the $r^{1/4}$ luminosity profile followed by
the elliptical galaxies and the bulges can be deprojected obtaining an
intrinsic density law.  In the last years, evidence has been presented
that the Young law can indeed be a poor fit, and the S\'ersic law,
$r^{1/n}$, was presented as a better-suited function (Caon et
al. 1993).  By analogy, the case of spiral-galaxy bulges has been
revisited and new trends have emerged, showing a large spread in $n$
peaking at $n\simeq 1$ (exponential bulges), but ranging up to
$n\simeq 15)$.  (Andredakis \& Sanders 1994; de Jong 1997; Andredakis
et al. 1995; Broeils \& Courteau 1997; Prugnel \& Simien 1997).  The
deprojection of the general $r^{1/n}$ law can then approximated by the
analytical formula (Ciotti 1991):

\begin{equation}
\label{eq:rho_unosuenne}
\rho_{\rm B}(r) \propto {{\rm e}^{-bx^{1/n}}\over x^{(2n-1)/2n}}
\end{equation}
where x=${r/R_{\rm eB}}$ and $b \approx 1.99865n-0.326968$, as given
by Caon et al. (1993), which is valid for every value of the flattening of 
the galaxy. $R_{\rm eB}$ is the bulge effective radius, corresponding
to the distance from center in which half of the total bulge
luminosity is contained.

For the disk component, the observed exponential profile can be deprojected
generating the law:
\begin{equation}
\label{eq:myrhod}
\rho(r)= \rho_{\rm 0D} \alpha K_0(\alpha r)
\end{equation}
where $K_0$ is the incomplete Bessel function of order zero, and
$1/\alpha$ is the disk scale length. The disk effective radius is
$R_{\rm eD}$=1.6784/$\alpha$.
It is important here to note that the ``disk'' component we adopted is
not a real, infinitely flat disk, but a spheroid of finite thickness
and exponential profile.

The above laws may be cut at a radius $R_{\rm max}$ for computation
purposes, and put together to produce a very wide range of complex
global density distributions. The introduction of an (usually dark)
halo among the galaxy components is needed when kinematics at large
galactocentric radii are considered.  In this paper, we shall explore
structures composed by bright components only, omitting the
introduction of a dark halo.  This choice follows the fact that in the
most part of stellar rotation curves presented in the literature the
regions explored are limited to the luminous component inside 
$1 \div 2 R_e$.  

\subsection{The velocity distribution}

When the spatial dependence of the distribution function $f(x,v)$ is
determined by the combination of the above density laws, the velocity
distribution can be obtained by integrating the collisionless
Boltzmann equation producing the moments equations. This will make the
galaxy model self-consistent, warranting its dynamical equilibrium,
such as in real systems.

In order to close the system of the moment equations, some additional 
assumptions are needed on the model, that may be obtained by imposing
some restrictions on the shape of the velocity ellipsoid.
The most widely used assumption in literature at this regard is 
to assume $\sigma_{RR}=\sigma_{zz}$ everywhere in the galaxy, and then to
choose some easy-to-handle functional form of the dependence of the
azimuthal velocity dispersion $\sigma_{\phi\phi}$ from the radius and
the radial velocity dispersion (see, e.g., Satoh 1980). 
However, this hypothesis implicitly describes a system with two
integrals of motion or $f(E, L_Z)$ (Binney \etal 1990) that has been
shown (Merritt 1999; Emsellem \etal 1999; Cretton \etal 1999) to fail
in correctly reproducing the observed behavior of many objects. 

In order to explore the three-integrals distribution functions $f(E,
L_z, I_3)$ one is forced to relax the $\sigma_{RR}=\sigma_{zz}$
assumption (see, e.g., Dehnen \& Gehrard 1993).  The simplest way to
do that is to assume $\sigma_{RR}=\sigma_{\phi\phi}$ everywhere in the
galaxy and that {\em the ratio} between the vertical and the radial
velocity dispersions is constant in the galaxy (thus including the
special case of isotropic models).

We point out that with this choice we are presenting just a special
set of anisotropic models involving a third integral. Nevertheless, it
is likely that the models can be used to derive a ``mean'' vertical
anisotropy for every observed galaxy, useful to derive general
relations in large set of galaxies. For a deeper insight of the
kinematic structure of a single anisotropic galaxy, a more complex
model is needed.

More in detail, for every different component, we assume the following
constraints:

\begin{itemize}

\item The velocity distribution is locally Gaussian. This do not imply 
a Gaussian {\em total} velocity distribution;  

\item The velocity ellipsoid's principal axes remain aligned with the 
cylindric coordinate system $(R, \phi, z)$, so that 
$\sigma_{Rz} =\sigma_{R\phi} = \sigma_{\phi z} = 0$; this assumption 
describes a situation similar to that of the solar neighborhood, where the 
deviation of the velocity ellipsoids from the cylindric coordinate axes is 
low.

\item The ratios of the components of the tensor $\sigma_{ij}^2$ are 
constant inside the galaxy, so that we can write
\begin{equation}
\tens{\sigma^2}= s(\vec{x}) \tens{\sigma}(0)
\end{equation} 
where $s(\vec{x})$ is a scalar function of the coordinates. The choice
produces velocity ellipsoids with same shape in every point of the
galaxy. 

\item A final requirement is the adoption of: 
$\sigma_{\phi\phi} = \sigma_{RR}$, in every region of the galaxy. 

\end{itemize}

Under these limits, needed to simplify the equations and to reduce the
number of free parameters, we can describe the velocity dispersion
field by means of $\sigma_{RR}(R,z)$ and the homogeneous {\em anisotropy
parameter} $\beta$, where
\begin{equation}
\beta \equiv 1 - \frac{\sigma^2_{zz}}{\sigma^2_{RR}} 
\end{equation}

Note that we do allow different components of the galaxy to have
different $\beta$ values. For instance, we can assume the disk
to have a velocity distribution more isotropic of the bulge, as well
as the contrary, and calculate the global line-of-sight velocity
distribution. Such differences are not only expected, but also 
at least hinted by the applications of models using only one single
value of anisotropy in a galaxy (\cite{cretton99}).

\section{The equations}

A second step is to derive the main galaxy parameters to be directly
compared with the observed photometric and kinematic data.

Independently from the chosen density distribution, the Jeans
equations in cylindrical coordinates for a steady-state, axisymmetric
system having a velocity ellipsoid described by the above equations
produce:

\begin{eqnarray}
\sigma_{RR}^2(R,z) &=& \frac{1}{\rho ( 1- \beta )} \int_z^\infty \rho 
\frac{ \partial \Phi }{\partial z} \, {\rm d}z \label{jeans2} \\
V_{\rm rot}^2(R,z) &=&  
\sigma_{RR}^2 \frac{\partial \ln \rho \sigma_{RR}^2(R,z) }{\partial \ln R} + 
 R \frac{\partial \Phi}{\partial R} 
\label{jeans1}
\end{eqnarray}

The partial derivatives of the potential in these equations can be calculated 
by making use of the relations for a single elliptical component and
of the additive property of the potential:

\begin{eqnarray}
\frac{ \partial \Phi}{\partial R} &=& \sum_i 2 \pi G R \sqrt{1-e_i^2} 
\int_0^\infty \frac{\rho_i(m^2)}{(\tau +1)^2
\sqrt{\tau+1-e_i^2}}{\rm d}\tau \label{deriv1}
\\
\frac{ \partial \Phi}{\partial z} &=& \sum_i 2 \pi G z \sqrt{1-e_i^2}
\int_0^\infty  \frac{\rho_i(m^2)}{(\tau+1)(\tau+1-e_i^2)^{3/2}}{\rm d}\tau
\label{deriv2}
\end{eqnarray}

We defined, following the notation of Binney \& Tremaine (1987),
\begin{equation}
m^2 \equiv \frac{R^2}{1+\tau} + \frac{z^2}{\tau + 1 - e^2}
\end{equation}
where $e = \sqrt{1-b/a}$ is the eccentricity of the component.  

In the regions where the bulge and disk luminosities are comparable,
the superposition of the rapid rotation of the disk to the slower
rotation of the other components produces in general a non-Gaussian,
2-peak line-of-sight velocity distribution (LOSVD).  To obtain the
projected $\{ I, V, \sigma \}$ for this distribution to be compared
with the data is not an easy task, since we must integrate the whole
LOSVD along the line-of-sight. However, the observed line-of-sight
velocity distribution is often parameterized in terms of $\{V, \sigma,
h_3, h_4\}$ of a Gauss-Hermite series( as described by van der Marel
\& Franx, 1993; see Appendix A).  We chose to evaluate the value of these
parameters, instead of deriving the full LOSVD from the unprojected
velocity distribution.

First, we have to correct the $V, \sigma$ unprojected values by
properly taking into account the effect of anisotropy and inclination
of the galaxy. The rotational velocity along the line of sight is $<V>
= V \sin i \cos \phi$, where $\phi$ is the azimuthal coordinate and
$i$ is the angle of inclination of the galaxy ($i=90^\circ$
corresponds to an edge-on galaxy); the component of the velocity
dispersion in the direction of the line of sight can be expressed as
\hbox{$\sigma^2 \left( 1 -\beta \cos^2 i \right)$} so that 
the second order projected moment is
 $<V^2> = V^2 \sin^2 i \cos^2 \phi + \sigma^2 \left( 1 -\beta \cos^2 i \right)$. 

We now integrate along the line of sight the 5 projected momenta $\{\rho, <V>,
<V^2>, <V^3>, <V^4>\}$. This can be done easily as long as we assume
that every component has a Gaussian local velocity distribution, by
making use of the general formula:
\begin{equation}
\label{eq:project}
<Q>(R_{\rm proj},z) = \frac{2 \int_{R_{\rm proj}}^{\sqrt{R_{\rm max}^2-z^2}} 
\rho(R,z) Q(R,z)R\,{\rm d}R}
{2 \int_{R_{\rm proj}}^{\sqrt{R_{\rm max}^2-z^2}} \rho(R,z) R\,{\rm d}R}
\end{equation}
where the notation $<Q>(R_{\rm proj},z)$ is used to indicate the
generic quantity $Q$ integrated along the line of sight at the
position $(R_{\rm proj},z)$ on the plane of the sky.  The values of
$\{V, \sigma, h_3, h_4\}$ cannot be obtained by a straightforward 
integration along the line of sight, because of their intrinsic
non-linearity; instead, we can integrate the velocity moments, which
are linear.  Also note that we do not restrict our investigation to
the case of major-or minor axis profiles.

Then we make use of the projected values of such momenta to calculate 
the skewness $\xi_3$ and of the kurtosis $\xi_4$ of the LOSVD at any 
given radius. The $\{V, \sigma, h_3, h_4\}$ values can now be obtained with the
help of the approximated formulas (van der Marel \& Franx 1993):
\begin{itemize}
\item $h_3= \xi_3/(4\sqrt{3})$
\item $h_4= (\xi_4-3)/(8\sqrt{6})$
\item $\sigma= (<V^2>_p-<V>_p^2)/(1+h_4\sqrt{6})$
\item $V= <V>_p-\sigma \cdot h_3 \cdot \sqrt{3}$
\end{itemize}

\section{ The parameters }

The model described in the previous sections is then determined by
$4n+1$ free parameters, where $n$ is the number of components of the
model: the total mass $M$, the scale length $R_e$, the eccentricity $e$
and the parameter $\beta$ defined above, plus the inclination angle of
the galaxy. In the simple case of a bulge+disk model, this lead to 9
free parameters.

Two of these parameters can be considered scale factors respectively
for the mass (and so, the kinematic quantities) and the lengths. 
We want to choose these scale factors in a way that makes simpler the
comparison with the observations, so we select the {\em global} mass
$M_{\rm tot}$ of the model and the {\em global} $\re$ (defined as the
major axis isophotal radius inside which lies half of the total
luminosity) as the scale factors.

The models are then presented in a dimensionless form, having, in the
bulge+disk case, seven (dimensionless) parameters: the two axial
ratios, the ratios $\eta= M_{\rm B}/M_{\rm D}$ and $\xi=R_{\rm
eB}/R_{\rm eD}$, the two values of $\beta_B$ and $\beta_D$ and the
ratio of the $(M/L)$ values. The parameter $\eta$ represents the
`physical' bulge-to-disk ratio and the $\xi$ the concentration of the
bulge with respect to the disk. In the general, $n-$component model, 
we would face $5n-3$ parameters.

Not every region of the parameters space is interesting; on the one
hand, as will be shown in next section, for some combination of the
parameters wide regions of the galaxy model cannot reach dynamical
equilibrium.  On the other hand, some of these parameters are never
observed in real galaxies: in particular, we kept fixed a $(b/a)_{\rm
D} = 0.2$ value, which seems a reasonable value for the observed
disks, and we restricted ourselves to the case where the $M/L$ ratio
of the bulge component is greater of the corresponding disk value.

In the simplest case of a bulge+disk model, we do expect that the four
remaining free parameters will cover two different physical
characteristics: the couple $(\eta, \xi)$ will somehow reproduce the
morphological behavior of galaxies along the Hubble diagram, while the
couple $(e_{\rm B}, \beta)$ will give us informations about the
dynamic of the bulge ( rotation- or pressure- supported ) of the
galaxy.

\section{ The results}

The model offers several possible applications of the stellar
component of disc galaxies to the observed data.  We choose here to
illustrate two general applications: the deduction of the mass ratio
between the disk and the bulge and the possibility to detect the
presence of anisotropies in the velocity dispersion field. In the end,
we will discuss in detail the case of a well-known S0 galaxy, NGC
5866, for which both photometric and kinematic data are available.
 
\subsection{ The models and the Hubble diagram }

\begin{figure*}[t]
\centerline{\psfig{file=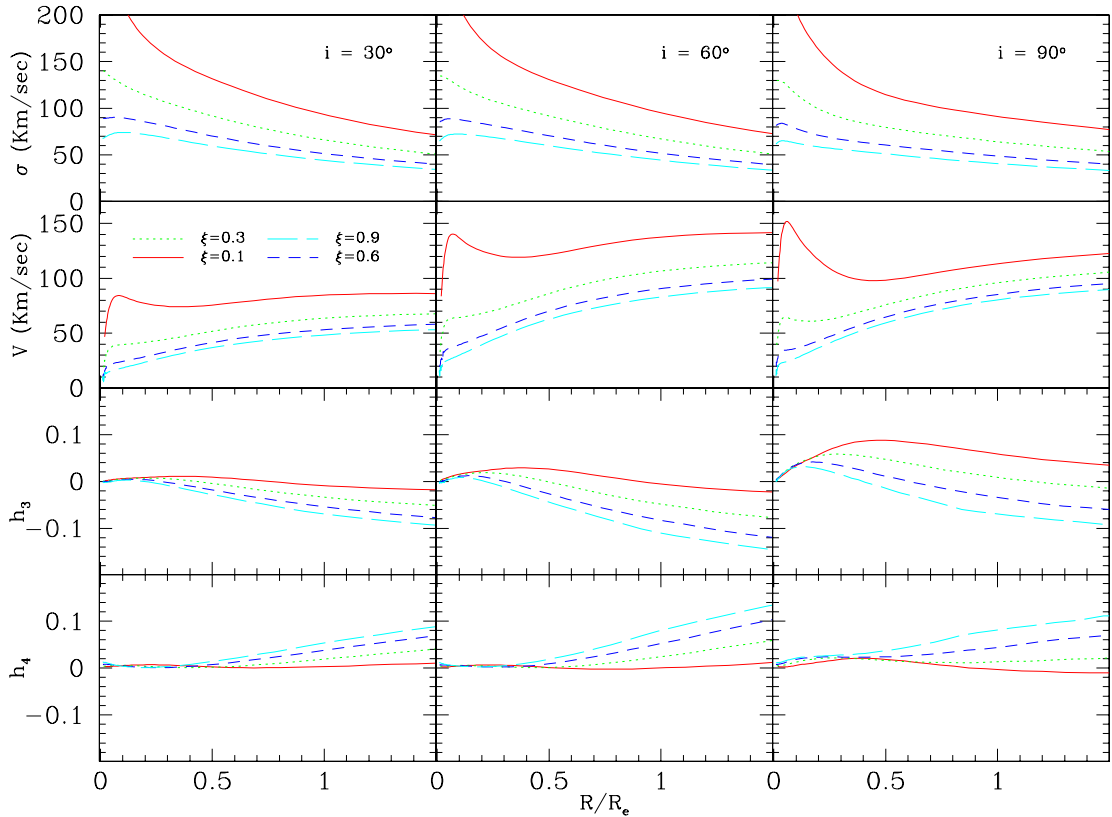,width=16cm,angle=270}}
\caption{ Rotation curves and velocity dispersion profiles for Sa galaxies,
varying the concentration of the bulge relative to the disk
components, and the inclination angle $i$. The two panels below also
show the $h_3$, $h_4$ parameters profiles as described by the model.
All the quantities are referred to a galaxy of $M=10^{11} \Msun$, and
$\re=10\,\kpc$. The scaling factor is described into the text. }
\label{fig:Saang}
\end{figure*}

The classification of galaxies in a single unificating scheme has been
one of the earlier results in the study of stellar systems.  The
details of this classification are, however, still under discussion.
Along the sequence of disk galaxies, the classification of galaxies
may be expressed in terms of photometric properties, such as the
relative weights of the bulge and disk component and their relative
concentration. Our model allow a dynamical deduction of these
properties.  We can compare observed velocity- and velocity
dispersion- curves with that of a model galaxy, projected at the same
viewing angle of the real galaxy and integrated along the
line-of-sight. In such a way, one can deduce the disk-to-bulge ratio
of the galaxy in terms of mass ratio and relative concentration of the
two components. We expect that this dynamical description may give
different results with respect to the disk/bulge ratio photometrically
deduced, and allow to correct the $M/L$ ratio
assumed and eventually detect the presence of dark matter contained in
the inner/intermediate portion of the galaxy.

To make this comparison easy, the dimensionless quantities $V_{\rm
mod}$, $\sigma_{\rm mod}$, produced by the model should be correlated
with the physical parameters. Since the model quantities are produced
in the dimensionless system having $GM_{\rm tot}=1$ and $R_{\rm e,tot}
=1$, they can be related to the observables by the use of the simple
scaling law:
\begin{eqnarray*}
V_{\rm obs} &=& \sqrt{\frac{GM_{\rm tot}}{R_{\rm e,tot}}}\cdot V_{\rm mod} \\ 
\sigma_{\rm obs} &=& \sqrt{\frac{GM_{\rm tot}}{R_{\rm e,tot}}}\cdot 
\sigma_{\rm mod}
\end{eqnarray*}  

We then produced a set of rotation- and velocity dispersion curves for
six type of galaxies having different bulge-to-disk mass ratio
($\eta$) and concentrations ($\xi$), and with a given $M_{\rm
tot}=10^{11} \, \Msun$, and $R_{\rm e,tot}=10\, \kpc$. In this
simulation, all the components have an isotropic velocity dispersion.
In order to limit the number of free parameters, we stick to the
classical case of a $r^{1/4}$ bulge embedded in an exponential
disk. The $M/L$ ratio of the bulge is assumed to be twice the
corresponding value of the disk.  In the next paragraph we will show 
the relevance of an anisotropic velocity dispersion in modifying the 
observed curves.

The curves were calculated for the galaxy seen edge-on ($90^\circ$)
and at an inclination of $60^\circ$ and $30^\circ$.  The choice of the
$\eta$ values has been made trying to reproduce the main Hubble
stages, from S0 to Sd, in agreement with the values deduced by Simien
\& de Vaucouleurs (1986). The whole grid of calculation is available
on request in electronic form. We are able to reproduce the whole
sequence assuming different slopes or different luminosity laws for
the bulge and the disk but of course this profusion of data cannot be
presented here.

In Fig.~\ref{fig:Saang} we show the kinematic curves for the case of
Sa galaxies (here simply defined as galaxies having an $\eta=0.68$).
They are arranged with the curves having different $\xi$ shown in the
same panel.

The relation between the observed curves with that presented in the figures 
is then:

\begin{eqnarray*}
V_{\rm obs} &=& \sqrt{\frac{M_{\rm tot}/[ 10^{11} \, \Msun ]}
{R_{\rm e,tot}/[ 10\, \hbox{kpc}]}}\cdot V_{\rm fig} \\ 
\sigma_{\rm obs} &=& \sqrt{\frac{M_{\rm tot}/[ 10^{11} \, \Msun]}
{R_{\rm e,tot}/[ 10\, \hbox{kpc}]}}\cdot 
\sigma_{\rm fig}
\end{eqnarray*}  
with masses computed in solar mass units and radii in Kpc.

This procedure allows, within some limits, to consider the galaxy as a
unique structure, with a typical bulge/disk ratio and bulge
concentration. However, it fits also into the idea that it is possible
to distinguish dynamically the galaxy type. This approach has been
faced by Persic and Salucci (1996) on the basis of a sample of
galaxies taken from the literature. Working on the superposition of
nearly 1000 observed gas rotation curves grouped for Hubble type, the
authors derived a circular rotation curve for every given Hubble type.
However, since this work was mainly aimed at the detection of dark
matter, more care is devoted to the outer part of the rotation curves;
in addition, a poor statistic is available on the early-type disk
galaxies S0 and Sa.

\subsection{ The anisotropy in bulges }

The low rotational velocities found in elliptical galaxies with
respect to their flattening (Bertola \& Capaccioli 1977, Kormendy \&
Illingworth 1977) can be interpreted in terms of anisotropy in the stellar 
velocity distribution (Binney 1978). The same indications arose from
the study of the velocity dispersion profiles (Tonry 1983).  For
similarity with elliptical galaxies, we can expect that some
anisotropy could be present in bulge dynamics also. On the contrary,
the presence of anisotropy in the disk component of the galaxies
should be less relevant, due to the lower value of the velocity
dispersion of disks compared with bulges or elliptical galaxies.

The study of velocity anisotropy in disk galaxies may give indications
on the differences existing between bulges and ellipticals,
contributing to understand the formation process of this two classes
of objects.  However, due to the composite nature of disk galaxies, to
measure the anisotropy of the velocity distribution from the
observables $(V,\sigma)$ is much more difficult than in the case of
elliptical galaxies.  Working on the same set of data, concerning four
S0 and Sa galaxies, Kormendy \& Illingworth (1982) deduced the isotropy
of their bulges, while Whitmore, Rubin and Ford (1985) reached the
opposite conclusion.  Both authors used the $V/\sigma$ test, first
proposed by Binney (1976) for Elliptical galaxies. The fact that the
same data may be interpreted in such different ways clearly
demonstrates the ambiguity of this test, when applied to the bulges of
disk galaxies. 

In a model of galaxy, the shape of the potential defines, on the galaxy
plane, the shape of the rotation- and velocity dispersion- profiles.
However, the ratio of the central velocity dispersion to the maximum
rotational velocity is dependent both from the eccentricity $e$ and
from the anisotropy parameter $\beta$: one may obtain in a model a
higher $V/\sigma$ ratio increasing the eccentricity and/or decreasing
the anisotropy parameter. As a reverse argument,we can deduce the
galaxy potential by fitting the observed rotation curve along the
apparent major axis but we cannot deduce from this curve the values of
$e$ or $\beta$. To define the values of these parameters, we need more
observable constraints.

One of this may arose from the photometry; in particular, from the
value of the axial ratio obtained by fitting the observed isophotes. In such
a case, the $\beta$ parameter can be, in principle, derived from the
major axis kinematic curves ($V$ and $\sigma$) only. Such a case is
however so rarely found in the literature, due to the great errors
that can derive from the decomposition algorithms, that the
photometric methods cannot be considered of general application
(Seyfert \& Scorza 1996).

A different approach may allow to deduce the ellipticity and anisotropy
with a simultaneous fit of several rotation curves taken at different
position angles or with offsets parallel to the major axis. This kind
of data is available in the literature for an increasing number of
galaxies and may be used to derive the anisotropy in their velocity
distribution.The model presented here can support this kind of
approach to estimate the amount of anisotropy present in an observed
galaxy. 

In the following, we shall discuss two points: first, how the
anisotropy is involved in the stability of a galaxy; second, how much
our model is able to distinguish between different anisotropies by
means of the rotation- and velocity dispersion- curves.

\begin{figure}[h]
\centerline{\psfig{file=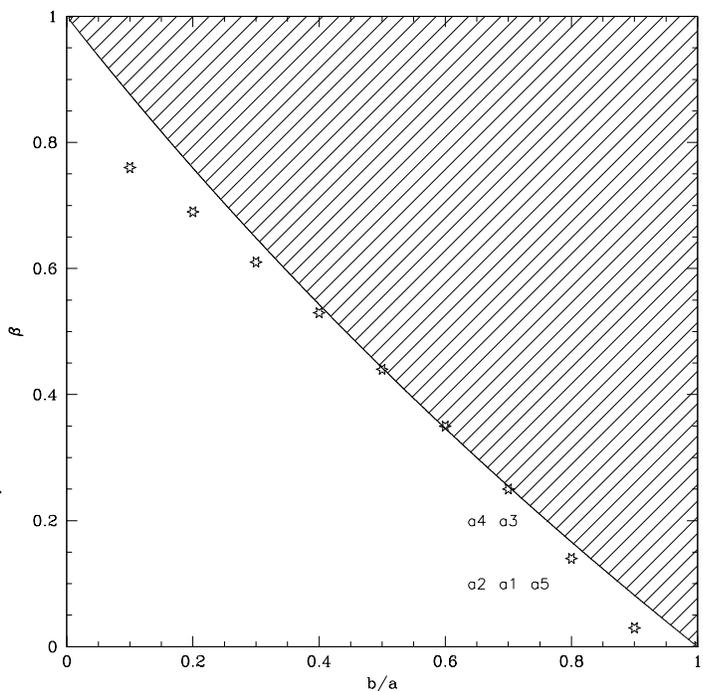,width=10cm}}
\caption{ Axial ratios and anisotropy of our models.
Below the solid line lies the physically allowed space of parameters,
in the simple case of a single homogeneous spheroid, and under our
hypothesis of constant anisotropy.  The maximum $\beta$ allowed for
different values of the axial ratio in the more realistic case of
two-components galaxies are shown as starred points.  The labels show
the location of the models described in
Fig.~\protect\ref{fig:anisotropi}. }
\label{fig:be}
\end{figure}

\begin{figure*}[t]
\centerline{\psfig{file=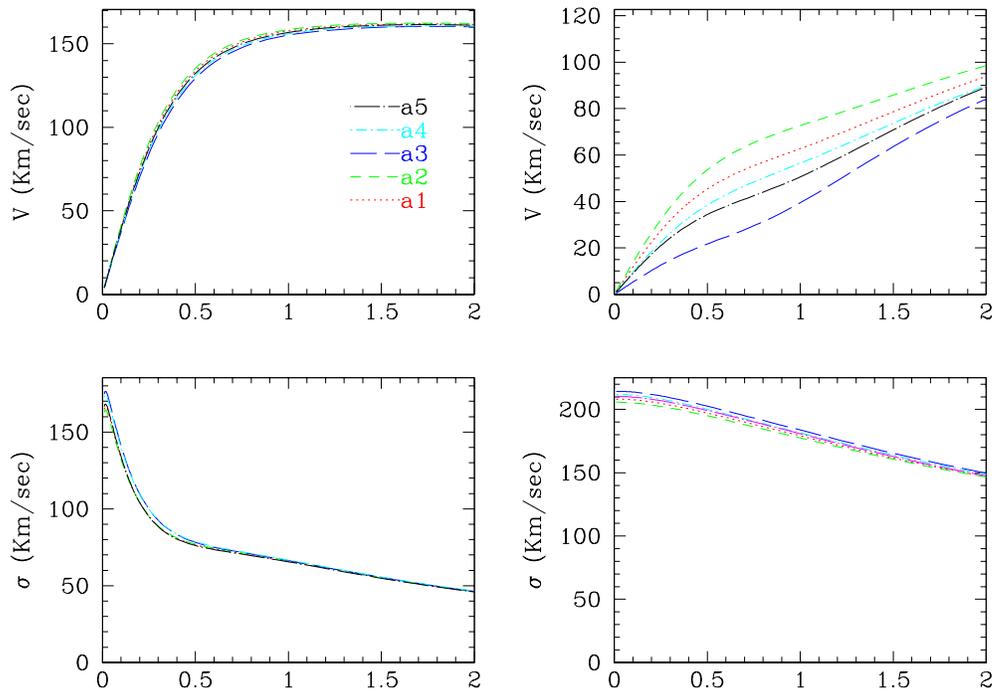,width=14cm,angle=270}}
\caption{ An example of different kinematic curves along the major axis 
(left panels) and along an offset parallel to the major axis but
shifted by $0.5\, R_e$.  The different curves correspond to an S0
galaxy with fixed $\xi, \eta$ values but different $(\beta, e)$. The
label of the different curves refers to the pairs of $(\beta, e)$
plotted in Fig.~\protect\ref{fig:be}. }
\label{fig:anisotropi}
\end{figure*}

\subsubsection{ Anisotropy and stability }

An instability mechanism may be generated by the local excess of
kinetic energy going in disordered motions with respect to the local
gravitational potential. Its genesis is the following:

Under the hypothesis that the only ordered motion in a galaxy be the
rotation around the $z$ axis, the $z$ component of the gravitation
force can only be equilibred by the $z$ component of the velocity
dispersion $\sigma_{zz}$, which is univocally determined by the
distribution of matter and {\em is not dependent from the anisotropy
parameter $\beta$ }. In such a galaxy, the {\em radial} component of
the velocity dispersion is higher as the anisotropy increases,
enhancing the importance of the asymmetric drift.  As $\beta$
approaches unity, the value of $\sigma_{RR}$ tends to infinity, and so
the asymmetric drift. Following this trend, the asymmetric drift
correction to the dynamical equation may overcome the gravitational
term of equation (\ref{jeans1}), pulling the galaxy outward and
destroying the dynamical equilibrium. As a consequence, once fixed in a
galaxy the axial ratio of the two components, not every values of
$\beta$ are allowed. The ``physical'' range of this parameter is
restricted to $0\le \beta\le \beta_{\rm max}(e)$, with $\beta_{\rm
max}(e)$ derived by imposing:
\begin{equation}
\Delta V^2_{\rm asymdrift} < V_{\rm c}^2
\label{eq:betae}
\end{equation}
in every point of the galaxy. 

Since the circular velocity is not dependent from the anisotropy while 
the correction for asymmetric drift is related to it only by means of a scale
factor, Eq.~(\ref{eq:betae}) can be rewritten as 
\begin{equation}
f(e) \equiv  \frac{\Delta V^2}{V^2_{\rm c}}(\beta=0) < 1- \beta
\end{equation}
so that the dependence from the $\beta$ parameter is easy to handle.

The $f(e)$ can be computed analytically in the limit case of
homogeneous spheroids only; the result is shown in Fig.~\ref{fig:be}
with a solid line.  In the realistic case, with two-components and
peaked density profiles, $f(e)$ must be computed numerically. We
estimated the limiting $\beta_{\rm max}$ from a set of models with
bulges of different ellipticities. The results are shown in
Fig.~\ref{fig:be} (starred symbols), showing that the homogeneous
spheroids curve can be used as a starting point for the analysis of
the realistic cases.

Again, we must point out here that this limiting values of $\beta$ are
only valid under our hypothesis of $\sigma_{RR}==\sigma{\phi\phi}$,
and should certainly be recalculated in different kinematic
conditions.

\begin{table*}[t]
\begin{center}
\caption[]{Parameters from the Dynamical Model for NGC 5866. The labels $b$ and $d$ refers
to the bulge and disk components.  }
\label{tab:ngc5866}
\begin{tabular}{cccccccccccccc}
\hline
\noalign{\smallskip}
\multicolumn{1}{c}{} 		 	&
\multicolumn{2}{c}{scale radius} 	&
\multicolumn{1}{c}{} 		 	&
\multicolumn{2}{c}{axial ratio}  	&
\multicolumn{1}{c}{} 		 	&
\multicolumn{2}{c}{Luminosity} 		&
\multicolumn{1}{c}{} 		 	&
\multicolumn{2}{c}{Mass [10$^{10} M_\odot$]}&
\multicolumn{1}{c}{} 		 	&
\multicolumn{1}{c}{} 		 	
					\\ 
\cline{2-3} \cline{5-6}  \cline{8-9} \cline{11-12} \\
\vspace*{-.5cm}			\\
\multicolumn{1}{c}{} 		 	&
\multicolumn{1}{c}{r$_b$} 		&
\multicolumn{1}{c}{r$_d$} 		&
\multicolumn{1}{c}{} 		 	&
\multicolumn{1}{c}{(b/a)$_b$} 		&
\multicolumn{1}{c}{(b/a)$_d$} 		&
\multicolumn{1}{c}{} 		 	&
\multicolumn{1}{c}{L$_b$} 		&
\multicolumn{1}{c}{L$_d$} 		&
\multicolumn{1}{c}{} 		 	&
\multicolumn{1}{c}{M$_b$} 		&
\multicolumn{1}{c}{M$_d$} 		&
\multicolumn{1}{c}{} 		 	&
\multicolumn{1}{c}{i} 
					\\
\noalign{\smallskip}
\hline
\noalign{\smallskip}
&$35\prime\prime$ &
$ 15\prime\prime$ &&
0.8 & 0.15 &&
62 \% & 38 \% &&
5.58 & 1.72 && $71^\circ$ \\
\hline
\end{tabular} 
\end{center}
\end{table*}

\subsubsection{ Inferring the anisotropy from observations. }

An observer may estimate the anisotropy present in a galaxy only if
the difference between models with different anisotropy are greater
than the observational errors. In order to give a rough estimate of
the diagnostic power of the model in determining the anisotropy
parameter $\beta$, we produced several models in different regions of
the diagram of Fig.~\ref{fig:be}. Looking at the literature, a
conservative estimate for the observational errors in kinematic
curves may be established assuming $\Delta V=20$ km/sec and $\Delta
\sigma$ = 40 Km/s.

Curves of a typical model with different anisotropy and bulge
flattening are shown in Fig.~\ref{fig:anisotropi}. Looking at this
figure, one can see that the difference in velocities generated by a
change of anisotropy of 0.1 produces a negligible variation along the
apparent galaxy major axis. On the contrary, significant variations
are observable along offset axes, e.g. setting the slit parallel to
the major axis but shifted by 0.5 $R_e$, as shown in the same
figure. The highest differences are visible in the rotation curve (as
great as 40 \kms) while the velocity dispersion curves are less
sensitive ($\le 10\kms$).

The sensitivity of the projected kinematic curves to variations of
$\beta$ or $e$ is different for galaxies lying in different regions of the 
$(\beta, e)$ plane. We can distinguish three typical cases:

\begin{enumerate}
\item {\bf Almost spherical bulges} 
{\em Under our assumption of homogeneous anisotropy}, an almost spherical
model (with bulge having $b/a>0.85$ ) does not allow stable solutions
with significant anisotropy. 

\item {\bf Strongly flattened bulges} 
This is the opposite situation ($b/a<0.4$).  In this case, the
diagnostic power of the model in determining anisotropy is limited.
The offset curves give only poor additional information with respect
to that obtained from the major axis data. In real galaxies this mean
flattening is observed only in the late-type spirals, where the disc
is preponderant with respect to the spheroidal component. As a
consequence, the bulge gives a low contribution to the global
kinematics.

\item{\bf Intermediate models}
The higher efficiency of the models in distinguishing the anisotropy
of a real galaxy from observations is found for galaxies with bulges
having intermediate flattening ($0.4\le b/a \le 0.85$).  The offset
rotation curve is very sensitive to the $\beta$ parameter changes, 
as described above in the example of Fig.~\ref{fig:anisotropi}.
The presence of an offset spectrum may delete, in this case, 
the uncertainties on the value of $\beta$ derived from observations
along the major axis. 

\end{enumerate}

\section{A test case: NGC~5866}

\begin{figure}[h]
\centerline{\psfig{figure=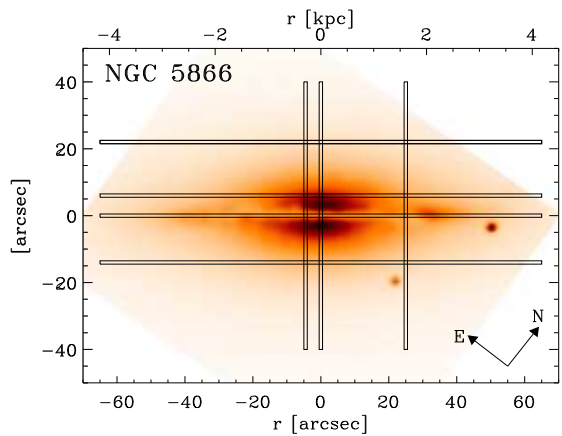,angle=90,width=11cm}}
\caption{ An image of the galaxy NGC 5866. The approximate position and 
extension of the kinematic profiles (Fisher 1997, Kormendy \&
Illingworth 1982) are shown superimposed on the multi-color image of
the galaxy (Peletier and Balcell 1997). }
\label{fig:slit}
\end{figure}

\begin{figure*}[t]
\centerline{\psfig{figure=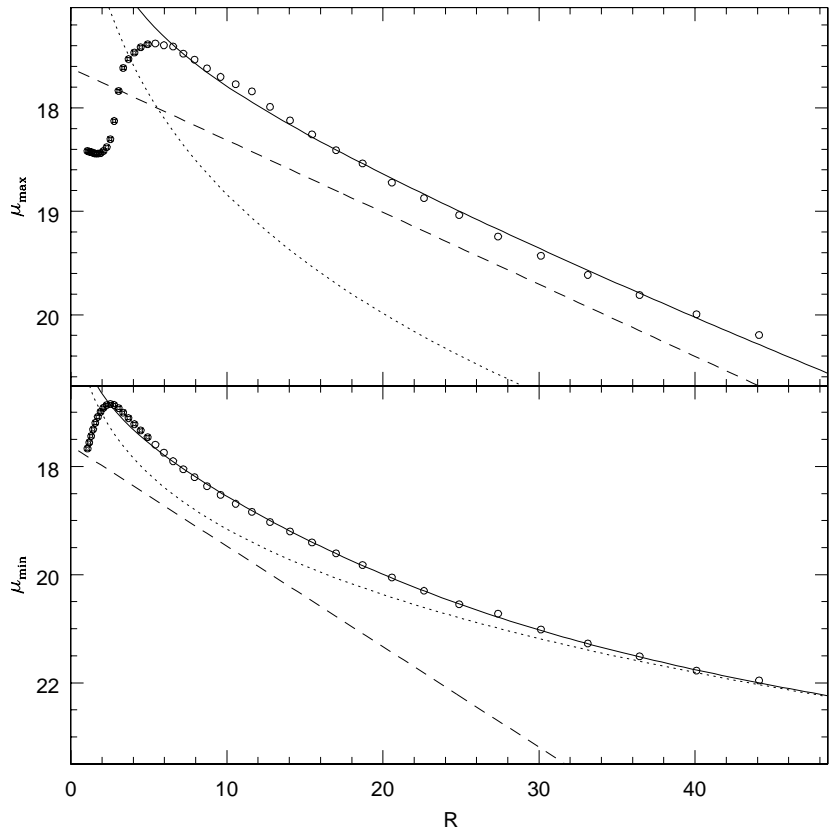,width=12cm}}
\caption{ Major and minor axis R-band photometric profiles, compared with 
the two components best-fit model. The dashed lines shown the contribution of 
the bulge and disk to the total surface brightness profiles (solid line). 
The crossed circle represent the points which have been excluded 
from the automatic fit due to the effect of the absorption of the dust lane.}
\label{fig:photo}
\end{figure*}

We tested our model using a real case of a galaxy for which a complete
set of photometric and kinematic data exists in the literature. We
chose NGC 5866, for which rotational velocity and velocity
dispersion profiles are available along the major axis and at
different offsets from the nucleus (Kormendy \& Illingworth
1982). This galaxy is an S0 seen almost edge-on, showing a prominent
dust lane on the major axis, clearly visible in Fig.~\ref{fig:slit}
and in the photometric profiles (Fig.~\ref{fig:photo}).

We took the recent kinematic data published by Fisher (1997) as well
as that published by Kormendy and Illingworth (1982).We adopted the
photometry made by Peletier and Balcell (1997). The full set of data
to be reproduced by our model include the profiles of velocity
dispersion and rotational velocity taken along 7 different axis across
the galaxy. The position of the
axis where kinematic data are available are shown in
Fig.~\ref{fig:slit}. The $h_3$ and $h_4$ parameters are also published 
but not for all the positions of the slit. 

The effect of the dust is marked in the innermost 5 arcsec along the
major axis.  For this reason we neglected the regions more heavily
obscured by the dust from our photometric fit (see
Fig.~\ref{fig:photo}), relaxing the fit along the major axis
luminosity profile, where the dust absorption is more
effective. 

\begin{figure*}[t]
\centerline{\psfig{figure=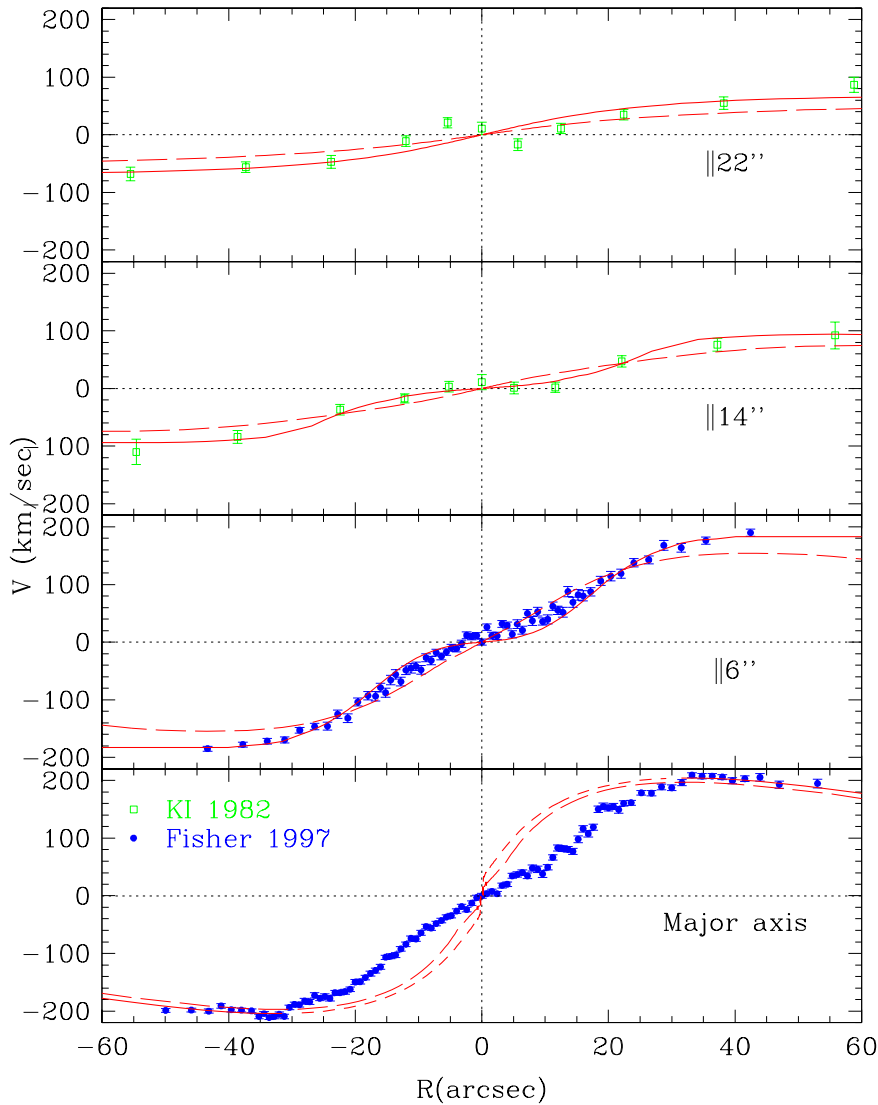,width=12cm}}
\caption{ Rotational velocity profile of stars and gas as compared to the model of the stellar kinematics, for the axis parallel to the major axis. The dashed line on the model correspond to the region obscured by the dust lane. In this plot we also show for comparison the best results for an anisotropic model with $\beta=0.1$ (long-dashed line).}
\label{fig:vmaj}
\end{figure*}

\begin{figure*}[t]
\centerline{\psfig{figure=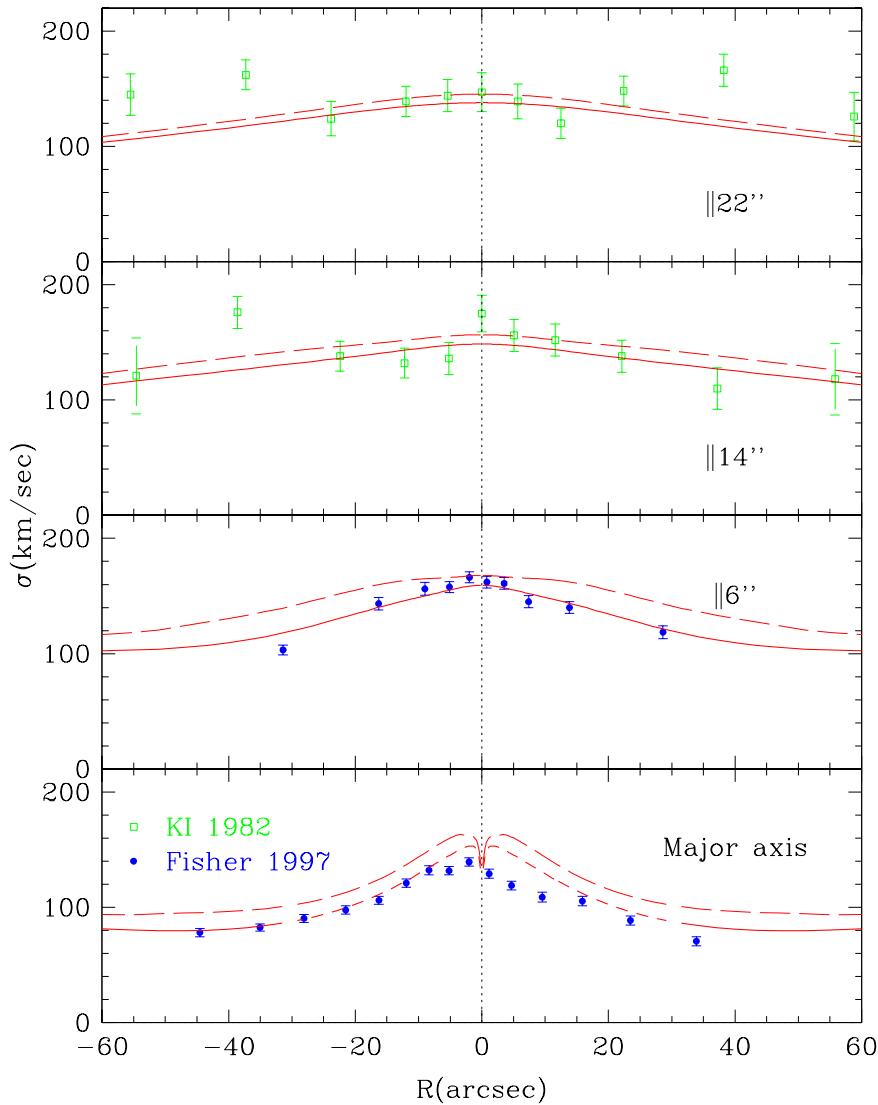,width=12cm}}
\caption{ Same as Fig.~\protect\ref{fig:vmaj}, but for the velocity dispersion profiles. Note the bad performance of the anisotropic model.}
\label{fig:sigmaj}
\end{figure*}

\begin{figure*}[t]
\centerline{\psfig{figure=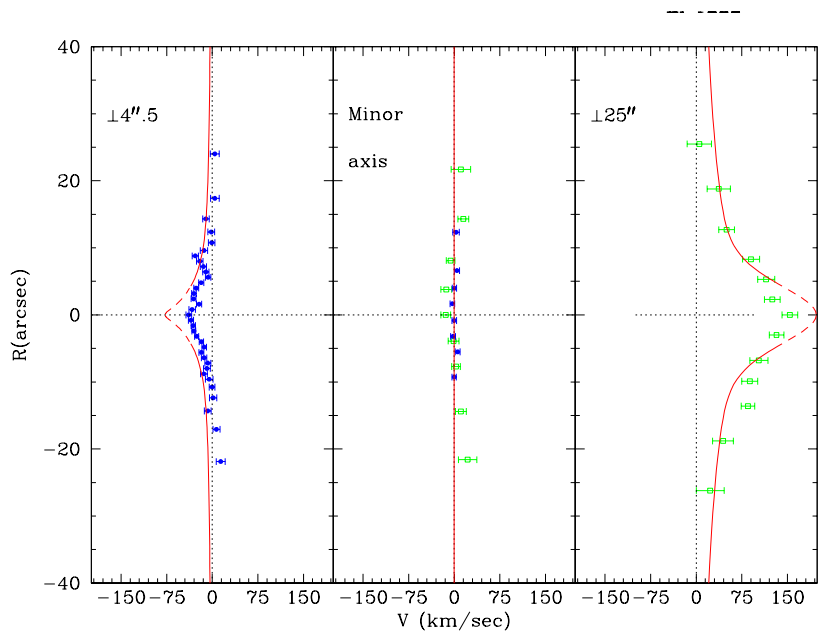,width=11.5cm,angle=270}}
\caption{ Same as in Fig.~\protect\ref{fig:vmaj}, but for the perpendicular cut shown in Fig.~\protect\ref{fig:slit}. We avoided to display the anisotropic model here. }
\label{fig:vmin}
\end{figure*}

\begin{figure*}[t]
\centerline{\psfig{figure=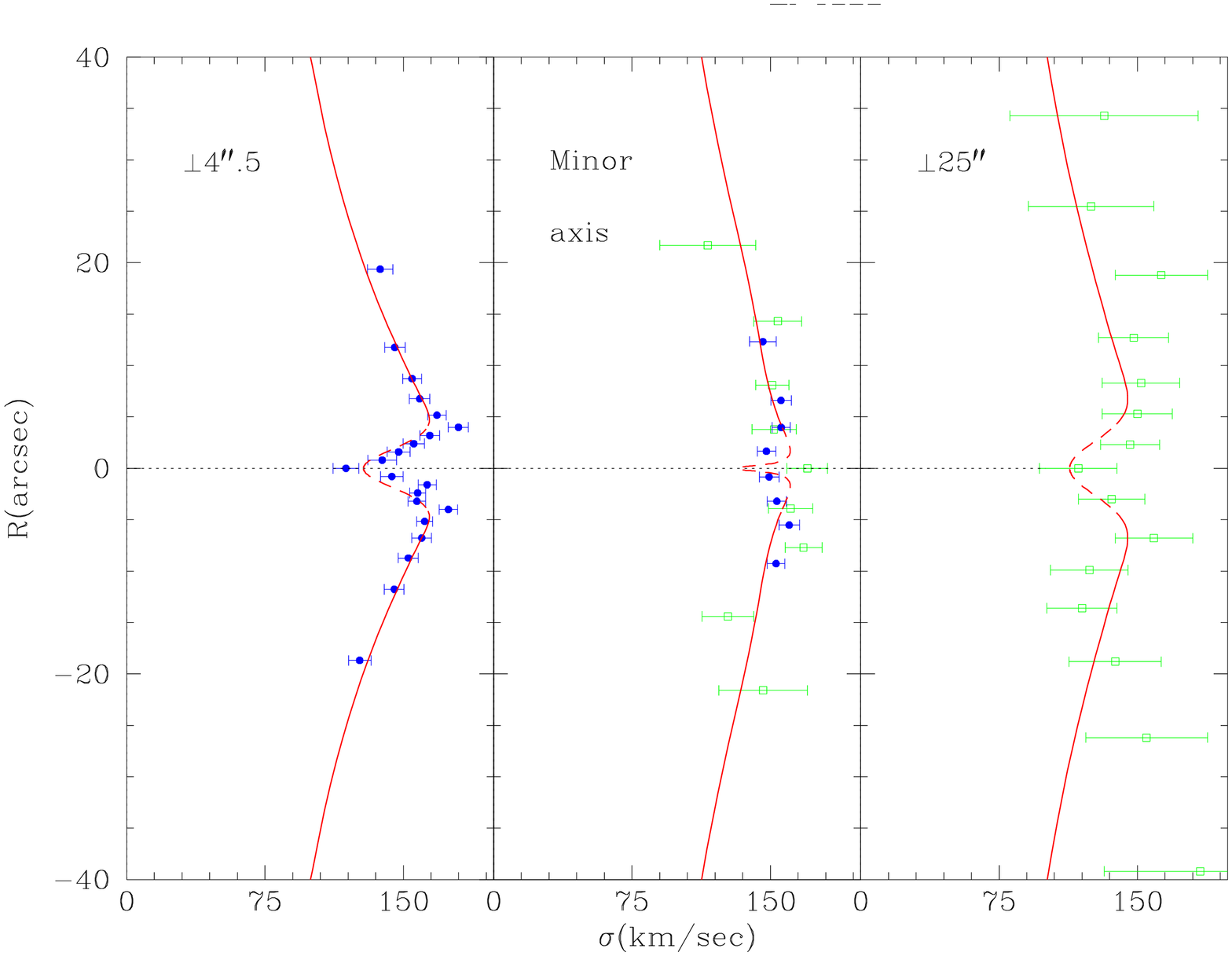,width=11.5cm,angle=270}}
\caption{ Same as in Fig.~\protect\ref{fig:sigmaj}, but for the perpendicular cut shown 
in Fig.~\protect\ref{fig:slit}.  }
\label{fig:sigmin}
\end{figure*}

The best-fit we adopted arise from a two-component (bulge+disk) model
whose parameters are shown in Tab.~\ref{tab:ngc5866}. With these values,
we computed by means of our model the expected, self-consistent
kinematics of this galaxy. According to our expectations, the results
reproduce most of the observed feature of the kinematics:

\begin{itemize} 

\item In general, the trend of the rotation curves and of the velocity 
dispersion profiles agrees everywhere with the model, {\em except in
the regions obscured by the dust} (see Fig.~\ref{fig:vmaj}).  The
explanation of this effect may be understood if the optical thickness
is high. In such a case, we only observe the stars on the outer edge
of the disk, whose line-of-sight velocity components are very low.

\item On the other hand, the presence of the dust is only slightly 
affecting the velocity dispersion, as expected in a dynamically hot
system such as an early-type galaxy. Our profiles agree with the
observations even in the regions obscured by dust. Along the major
axis the velocity dispersion profile is monotonically decreasing from
a peak value of 140 \kms in the central region to $\sigma \approx
74$ \kms at $40''$. On the major-axis offset positions, slightly
higher velocity dispersion are observed. The minor-axis data indicate
that the velocity dispersion rises marginally with increasing $z$
height (from $\approx 150$ km s$^-1$ to $\approx 158$ \kms at $6''$)
and decrease at larger radius. The same behavior is present in the
minor-axis offset spectra.  Fisher (1997) proposed that this is an
observational effect due to the dust obscuration: our model seem to
indicate that this is indeed a real effect, due to the dominance of
the cold disk component at $z \approx 0$.
 
\item Along the major axis, far from the region heavily obscured by the dust, 
the model rotation curves flattens according to the observed data.
Fitting the observed maximum rotational velocity of 203 \kms, we find
a total mass for the galaxy of $(7.3\pm 0.5) \cdot 10^{10}$ \Msun. 
Fairly rapid rotation is seen even at heights of several kpc 
above the disk plane.  At $z =6'', r= 40''$ the 
offset spectrum has a speed of 184 \kms. At
$z=22''$, equivalent to an height of 2 kpc above the disc plane, the
rotation curve reaches $\sim$100 \kms, a value comparable with the
mean dispersion $\sigma \approx 150 $ \kms observed in the bulge. This
is compatible with the isotropic model we are proposing, while no
model with significant amount of anisotropy has been found to be
able to improve the fit of the data.

\item No rotation is seen along the minor axis ($V_{min} = 3 \pm 3 $ km/sec), 
so we can likely exclude misalignments between the different
components or a significant bulge triaxiality.

\item The fit of the $h_3$ and $h_4$ parameters are marginally 
satisfactory on the major 
axis, taking into account the effects of the dust; but are poorer on
the offset at $6^{\prime\prime}$. However, having considered the
deviations from the pure Gaussian shape of the line profile greatly
improved the quality of the $V, \sigma$ fit.

\end{itemize}

\section{Conclusions}

Our self-consistent, two-components model of disc galaxy may reproduce
the photometric and the kinematic properties of the disc galaxies of
different morphological type. In particular:

\begin{enumerate}

\item The model is able to produce rotation curves, velocity dispersion 
profiles and h$_3$, h$_4$ curves for disc galaxies from S0 to Sd,
describing the different models in terms of relative weight and the
concentration of the bulge with respect to the disc. A library of
model curves, projected at different angles of galaxy inclination with
respect to the sky plane has been generated. The comparison of these
models with observations of real galaxies needs only two scaling
factors: the total mass and the effective radius.

\item The anisotropy of the velocity field  has been studied as 
instability generator inside the galaxy. We showed that for each given
value of the bulge axial ratio a stable stellar system cannot possess
a value of anisotropy greater than a fixed value.  In the extreme case
of spherical systems, only isotropic distributions of velocities are
allowed, under our assumption of anisotropy homogeneous across the galaxy.

\item The models show a strong diagnostic power in detecting anisotropy in
the bulges of the early-type disc galaxies when spectra outside of the
apparent major axis are available. The best results are obtainable in
the case of bulges of intermediate ($0.4\le b/a \le 0.85$) flattening,
close to the values observed in real galaxies.  In this case, an
anisotropy change of 15\% may give variations even of 40-50\% in the offset
rotation curves, as shown in Fig.~\ref{fig:anisotropi}.

\item The application of the model to edge-on S0 galaxy NGC~5866 has been 
presented. It suggests an isotropic distribution of velocity and is
able to reproduce the global galaxy kinematics ($V$ and $\sigma$ profiles).

\end{enumerate}

\subsection*{Acknowledgments}
This work has been partially supported by the grant ``Astrofisica e
Fisica Cosmica'' Fondi 40\% of the Italian Ministry of University and
Scientific and Technologic Research (MURST).

\appendix

\section*{Appendix A. Gauss-Hermite expansion of the line shape.}
\label{appendix:h3h4}
\setcounter{equation}{0}
\renewcommand{\theequation}{A\arabic{equation}}

One should address the question how the deviations from the pure
Gaussian profile are best quantified.

Van der Marel \& Franx (1993) represented the line profile as a sum of
orthogonal functions in a {\it Gauss-Hermite} series. The expansion
naturally leads to two parameters describing deviations from a
Gaussian: a parameter $h_3$ describing asymmetric deviations, and a
parameter $h_4$ describing symmetric deviations.

The LOSVDs $f(v)$ can then be described by a Gaussian plus third- and
fourth-order {\it Gauss-Hermite} functions:

\begin{equation}
f(v)=I_0\exp (-y^2/2) (1+h_3{\cal H}_3(y)+h_4{\cal H}_4(y))
\label{eq:losvd}
\end{equation}
with $y=(v-v_{fit})/\sigma_{fit}$, and where
\begin{equation}
{\cal H}_3(y)=(2\sqrt{2}y^3-3\sqrt{2}y)/\sqrt{6}
\label{eq:H3}
\end{equation}

\begin{equation}
{\cal H}_4(y)=(4y^4-12y^2+3)/\sqrt{24}
\label{eq:H4}
\end{equation}
are the standard {\it Gauss-Hermite} polynomials, $u_i=exp(-y^2/2) \times
{\cal H}_i(y)$ are the {\it Gauss-Hermite} basis functions, and $h_3$ and
$h_4$ are their amplitudes. $I_0$ is a normalization constant.

%
%

\end{document}